\documentclass [12pt]  {article}
\begin{document}
\title{%
Introduction to Doubly Special Relativity}
\author{ J.\ Kowalski--Glikman\thanks{e-mail address
jurekk@ift.uni.wroc.pl}\\  \\ {\em Institute for Theoretical
Physics}\\ {\em University of Wroc\l{}aw}\\ {\em Pl.\ Maxa Borna 9}\\
{\em Pl--50-204 Wroc\l{}aw, Poland}} \maketitle
\begin{abstract}
In these notes, based on the lectures given at 40th Winter School on Theoretical Physics, I review some aspects of Doubly Special Relativity (DSR). In particular, I discuss relation between DSR and quantum gravity, the formal structure of DSR proposal based on $\kappa$-Poincar\'e algebra and non-commutative $\kappa$-Minkowski space-time, as well us some results and puzzles related to DSR phenomenology.
\end{abstract}
\clearpage
\section{Introduction}
 What is the fate of Lorentz symmetry at  Planck scale? This question was the main theme of the Winter School and, as the reader could see from the proceedings, there are many possible answers. Here I would like to describe one possibility, whose central postulate is that in spite of the fact that departures from  Special Relativity are introduced   at scales close to Planck scale,  one  keeps unchanged the central physical message of the theory of relativity, namely the equivalence of all (inertial) observers. This justifies the term {\em Relativity} in the title.

To be more specific, let us  start with the set of postulates of \index{Doubly Special Relativity} Doubly Special Relativity\footnote{Some authors prefer to use the name Deformed Special Relativity, fortunately leading to the same acronym} (I will use the acronym DSR in what follows) or Special Relativity with Two Observer Independent Scales, as proposed in \cite{Amelino-Camelia:2000ge}, \cite{Amelino-Camelia:2000mn} (see also \cite{Kowalski-Glikman:2001gp}, \cite{Bruno:2001mw}.) These postulates can be formulate as follows.

\begin{itemize}
\item  One assumes that the relativity principle holds, i.e., equivalence of all inertial observers in
the sense of Galilean Relativity and Special Relativity is postulated.

\item There are  {\em two} observer independent scales: one of velocity $c$,
identified with the speed of light\footnote{Some readers may be confused already at this point since it is  often claimed that DSR predicts dependence of the speed of massless particles on energy they carry, so that the speed of light is energy (and wavelength) dependent. Then the question arises to which speed this postulate refers to. As I will show below there are, arguably, good reasons to believe that in DSR the speed of light equals $1$, independently of the energy. }, and second of dimension of mass $\kappa$
(or length $\lambda=\kappa^{-1}$), identified with the Planck mass. Of course, it is assumed that in the limit $\kappa \rightarrow\infty$ DSR becomes the standard Special Relativity. This postulate is the reason for the term ``Doubly''. Since it turns out that the action of symmetry generators must be deformed in DSR, one may talk about ``Deformed Special Relativity''.
\end{itemize}

It is a quite nontrivial problem, though, how these postulates can be realized in practice,  given the fact that at the Planck scale we are to have to do with two scales of length and/or mass. Indeed, we know both from the theory and numerous experiments that in Special Relativity different observers do attribute different lengths and masses to the same measurements: as it is well known, we have to do with Lorentz-FitzGerald contraction and relativistic corrections to mass. How is it then possible to have at the same time relativity principle and the observer-independent scale of length or mass? It turns out that it is possible, but the price to pay is quite high, as one presumably must describe space-time in terms of non-commutative geometry, and to talk about space-time symmetries, one should use the language of quantum groups.

It should be noted also that as an immediate consequence of the postulates 
 DSR theory should possess (like Galilean and Special Relativity theories) a ten
dimensional group of symmetries, corresponding to rotations, boosts, and
translations, which however, as a
result of the presence of the second scale, cannot be just the linear Poincar\'e group . This immediately poses a problem. Namely, if we have a theory with observer independent scale of mass, then it follows
that it should be expected that the standard Special Relativistic Casimir $E^2 - p^2
= m^2$ is to be replaced by some nonlinear mass-shell relation, between energy and
three-momentum (which would involve the parameter $\kappa$\footnote{Note however that there exists a class of models of DSR, in which the dispersion relation between energy and momentum is not deformed (see below.)}.) Thus the second scale $\kappa$ must be encoded into the mass-shell condition so that it is kept invariant by symmetry transformations. But then it
follows that the speed of massless particles defined as $\partial E/\partial p$ would be dependent on the energy
they carry, which makes it hard to understand what would be the operational
meaning of the observer-independent speed of light. Below I will
suggest  ways out of this dilemma.

I should warn the reader that the construction of the theory of Doubly Special relativity is not  completed yet; in fact we do not even have a single DSR candidate, which would satisfy all the requirements of internal and conceptual self-consistency. Nevertheless during the last three years many results have been obtained, and for example we now control pretty well the one particle sector of the theory, both technically and conceptually. However, many problems remain, for example, we still do not understand the multi-particle sector of  DSR theory.

The structure of this notes corresponds to the structure of the lectures I gave at the Winter School. The next section corresponding to the first lecture is devoted to the questions whether and how DSR could emerge as an appropriate limit of quantum gravity. The complete answer to these questions is still unknown but we have some number of evidences suggesting that indeed DSR may be rooted in quantum gravity. The third section of these notes is devoted to describing techniques used in a particular, best developed approach to DSR, based on the so-called $\kappa$-Poincar\'e algebra and $\kappa$-Minkowski space-time. In section 4 I would like to describe main results obtained in the DSR framework, as well as bunch of open problems, mainly related to the multi-particle processes.

\section{DSR from quantum gravity?}

If the DSR idea is correct, it is quite natural to expect that Doubly Special Relativity emerges somehow as a limit of quantum gravity. It is rather clear why it must be so. In the standard Special Relativity we have only one scale, and there is no natural way in which another scale of mass and/or length could be introduced purely in  Special Relativistic setting. On the other hand, in quantum gravity we have, in addition to the velocity scale $c$,  three additional dimensionful constants, $G$, $\hbar$ (which I often set equal  1 in what follows), and (sometimes) the cosmological constant $\Lambda$. The immediate idea is that in the limiting procedure, in which the gravitational interactions as well as quantum effects become negligible, and the space-time becomes effectively flat (at least locally), some trace of these constants remains, giving rise to new observer-independent scale $\kappa$. In this section I will try to convince the reader that such scenario may indeed result from quantum theory of gravity. 

Usually we take for granted that the $G \rightarrow 0$, (and possibly $\Lambda\rightarrow 0$  if we start with non-zero cosmological constant) \index{cosmological constant} limit of (quantum) gravity is just the Minkowski space-time. But perhaps this is not correct, and we are forced to take the limit (especially in the case in which point particles are present) such that either

\begin{enumerate}
\item $\lim_{G, \Lambda \rightarrow 0}\, \sqrt{\frac{G}{\Lambda}} = \kappa^{-1} \neq 0$, or alternatively, 
\item  $\lim_{G, \hbar \rightarrow 0}\, \sqrt{\frac{\hbar}{G}} = \kappa \neq 0$. 
\end{enumerate}

It is not clear which of these scenarios (if any) is realized in Nature, but there are some  indirect evidences in favor of the claim that indeed it might be so. 

Let us try to investigate the first scenario following the ideas presented in \cite{Amelino-Camelia:2003xp}. To this end let us consider first the three-dimensional quantum gravity with positive cosmological constant $\Lambda$. Then it is well known  \cite{Nelson:ba} that the excitations  of $3d$ quantum gravity with cosmological constant transform under representations of the quantum deformed
deSitter algebra $SO_q(3,1)$,  with
$z=\ln q$ behaving in the limit of small\footnote{Since in $3d$, the dimension of the gravitational constant is $1/kg$, we write $G = \kappa^{-1}$.} $\Lambda \hbar^2 /\kappa^2$ as $z  \approx  \sqrt{\Lambda} \hbar / \kappa$.

I will not discuss at this point the notion of quantum deformed algebras (Hopf algebras) in much details (the book \cite{majidbook} would be a good references for the reader who wants to study this exciting branch of mathematics.) It will suffice to say that quantum algebras consist of several structures, the most important for our current purposes would be the universal enveloping algebra, which could be understand as an algebra of brackets among generators, which are equal to some analytic functions of them. Thus the quantum algebra is a generalization of a Lie algebra, and it is worth observing that the former reduces to the latter in an appropriate limit. Quantum algebras start playing an important role in various branches of theoretical physics; in particular, in some cases, they can play a role of relativistic symmetries in some field theoretical models (see an excellent, pedagogical exposition in \cite{Agostini:2003vg}.) In the case of quantum algebra \index{quantum algebra $SO_q(3,1)$} $SO_q(3,1)$ the algebraic part looks as follows (the parameter $z$ used below is related to $q$ by $z=\ln q$)
\begin{eqnarray}
&& [M_{2,3},M_{1,3}]  =  {1 \over z} \sinh(z M_{1,2}) \cosh(z M_{0,3})
\nonumber \\
&& [M_{2,3},M_{1,2}] = M_{1,3}
\nonumber \\
&& [M_{2,3},M_{0,3}] = M_{0,2}
\nonumber \\
&& [M_{2,3},M_{0,2}] =  {1 \over z} \sinh(z M_{0,3}) \cosh(z M_{1,2})
\nonumber \\
&& [M_{1,3},M_{1,2}] = - M_{2,3}
\nonumber \\
&& [M_{1,3},M_{0,3}] = M_{0,1}
\nonumber \\
&& [M_{1,3},M_{0,1}] =   {1 \over z} \sinh(z M_{0,3}) \cosh(z M_{1,2})
\nonumber \\
&& [M_{1,2},M_{0,2}] = - M_{0,1}
\nonumber \\
&& [M_{1,2},M_{0,1}] = M_{0,2}
\nonumber \\
&& [M_{0,3},M_{0,2}] = M_{2,3}
\nonumber \\
&& [M_{0,3},M_{0,1}] = M_{1,3}
\nonumber \\
&& [M_{0,2},M_{0,1}] =  {1 \over z} \sinh(z M_{1,2}) \cosh(z M_{0,3})\label{1.1}
\end{eqnarray}
Since this is our first encounter with quantum algebra let us pause for a moment to discuss its main features. First of all, we observe that on the right hand sides we do not have linear functions generators, as in the Lie algebra case, but  some (analytic) functions of them. However we still assume that the brackets are antisymmetric and that  Jacobi identity holds.\newline

{\bf Exercise 1}. Convince yourself by direct inspection that for the algebra (\ref{1.1}) Jacobi identities indeed hold.\newline

It follows from this observation that  contrary to the Lie algebras case, we are now entitled to use any analytic functions of the initial set of generators as a basis of the quantum algebra (in the Lie algebra case we can only take linear combinations of them.) It should be stressed already at this point that quantum algebras possess more structures than just the enveloping algebra structure (for more details see \cite{majidbook}); some of them will be relevant in what follows. Note that in the limit $z\rightarrow0$ the algebra (\ref{1.1}) becomes the standard algebra $SO(3,1)$, and this is the reason for using the term $SO_q(3,1)$.\newline

{\bf Exercise 2}. Denote by $M_{\mu\nu}^z$ the generators of the algebra (\ref{1.1}) and by $M_{\mu\nu}$ the generators of the standard $SO(3,1)$ algebra (obviously the equation $\lim_{z\rightarrow0}\, M_{\mu\nu}^z = M_{\mu\nu}$ should hold.) Find  explicit expressions for $M_{\mu\nu}^z$ as functions of $M_{\mu\nu}$ and $z$. (If this exercise happens to be too hard do that only up to the next-to-leading order in $z$.)\newline

The  $SO(3,1)$ Lie algebra is the three dimensional de Sitter algebra and it is well known how to obtain the three dimensional Poincar\'e algebra from it. First of all one has to single out the energy and momentum generators of right physical dimension (note that the generators $M_{\mu\nu}$ of (\ref{1.1}) are dimensionless): one identifies three-momenta $P_\mu \equiv (E, P_i)$ ($\mu=1,2,3$, $i=1,2$) as appropriately rescaled generators $M_{0,\mu}$ and then one takes the  In\"om\"u--Wigner contraction limit (see, for example, \cite{barutraczka}.) 

Let us try therefore to proceed in an analogous way and contract the algebra (\ref{1.1}). To this aim we must first rescale some of the generators by an appropriate scale, provided by combination of dimensionful constants present in definition of the parameter $z$
\begin{eqnarray}
&& E = \sqrt{\Lambda}\hbar\, M_{0,3}
\nonumber \\
&& P_i = \sqrt{\Lambda}\hbar\, M_{0,i}
\nonumber \\
&& M = M_{1,2}
\nonumber \\
&& N_i = M_{i,3}
\label{1.2}
\end{eqnarray}
Taking now into account the relation $z  \approx  \sqrt{\Lambda} \hbar / \kappa$, which holds for small $\Lambda$,   from 
$$
[M_{2,3},M_{1,3}]  =  {1 \over z} \sinh(z M_{1,2}) \cosh(z M_{0,3})
$$
 we find 
\begin{equation}\label{1.3}
[N_2,N_1]  
= {\kappa \over \hbar \sqrt{\Lambda}} \sinh(\hbar \sqrt{\Lambda}/\kappa M)
\cosh( E/\kappa)
\end{equation}
Similarly from
$$
[M_{0,2},M_{0,1}] =  {1 \over z} \sinh(z M_{1,2}) \cosh(z M_{0,3})
$$
we get
\begin{equation}\label{1.4}
[P_2,P_1] 
= {{\sqrt{\Lambda}\hbar \kappa}   }\, \sinh(\sqrt{\Lambda} \hbar / \kappa\, M)
\cosh( E/\kappa)
\end{equation}
Similar substitutions can be made in other commutators of (\ref{1.1}).
Now going to the contraction limit $\Lambda\rightarrow0$, while keeping $\kappa$ constant we obtain  the following algebra
\begin{eqnarray}
&& [N_i,N_j]  =  - M \epsilon_{ij}\, \cosh( E/\kappa)
\nonumber \\
&& [M,N_i] = \epsilon_{ij} N^j
\nonumber \\
&& [N_i,E] = P_i
\nonumber \\
&& [N_i,P_j] = \delta_{ij}\, {\kappa}\, \sinh( E/\kappa)
\nonumber \\
&& [M,P_i] = \epsilon_{ij} P^j
\nonumber \\
&& [E,P_i] = 0
\nonumber \\
&& [P_2,P_1] = 0
\label{1.5}
\end{eqnarray}
This algebra is called the three dimensional $\kappa$-Poincar\'e algebra \index{$\kappa$-Poincar\'e algebra} (in the standard basis.)

It turns out that this contracted algebra is again a quantum algebra, i.e., after the contraction all the additional structures of $SO_q(3,1)$ became the analogous structures of the new algebra (which is not obvious a priori because, in principle, it may happen that during the contraction procedure additional structures of the quantum algebra may become not well defined). This really nontrivial and remarkable result has been obtained in early nineties in \cite{Lukierski:1991pn}, \cite{Lukierski:1991ff}.

Let us pause for a moment here to make couple of comments. First of all, one easily sees that in the limit $\kappa\rightarrow\infty$ from the $\kappa$-Poincar\'e algebra algebra (\ref{1.5}) one gets the standard Poincar\'e algebra. Second, we see that in this algebra both the Lorentz and translation sectors are deformed. However, as I have been stressing already, in the case of quantum algebras one is  free to change the basis of generators in arbitrary, analytic way. It turns out that there exists such a change of basis that the Lorentz part of the algebra becomes classical (i.e., undeformed.) Such a basis, derived in \cite{kappaM}, is called bicrossproduct (because of the remarkable bicrossproduct structure of the full quantum algebra, see \cite{majidbook}), and the Doubly Special Relativity model (both in 3 and 4 dimensions) based on such an algebra is called DSR1. \index{DSR1} In this basis the algebra looks as follows
\begin{eqnarray}
&& [N_i,N_j]  = -\epsilon_{ij}\, M
\nonumber \\
&& [M,N_i] = \epsilon_{ij} N^j
\nonumber \\
&& [N_i,E] = P_i
\nonumber  \\
&& [N_i,P_j] =   \delta_{ij}\, {\kappa\over 2}
 \left(  
 1 -e^{-2{E/ \kappa}}
 + {\vec{P}\,{}^{ 2}\over \kappa^2}  \right) - \, {1\over \kappa}
P_{i}P_{j}
\nonumber \\
&& [M,P_i] = \epsilon_{ij} P^j
\nonumber \\
&& [E,P_i] = 0
\nonumber \\
&& [P_1,P_2] = 0 ~.
\label{1.6}
\end{eqnarray}

{\bf Exercise 3}. Derive explicit transformations from variables in (\ref{1.5}) to variables in (\ref{1.6}) (solution can be found in \cite{kappaM}.)\newline

Note now that the algebra  (\ref{1.6}) is exactly of the form needed in Doubly Special Relativity. \index{Doubly Special Relativity} By construction this is the algebra of symmetries of flat space, being an appropriate limit of the algebra of symmetries of states of quantum gravity. Moreover it manifestly contains the observer-independent scale of dimension of mass $\kappa$. \newline

{\bf Exercise 4}. Check that $\kappa$ is the observer-independent scale in the  sense that if $|\vec{P}| = \kappa$, then $\delta|\vec{P}| =0$, where $\delta$ denotes the change under infinitesimal action of boosts (solution can be found in \cite{Kowalski-Glikman:2001gp}.)\newline.

This shows that, at least in principle, one can try to construct a theory, which satisfies principles of DSR, and that such a theory may be neither inconsistent, nor trivial. Of course to construct a theory of particle kinematics, with symmetries defined by (\ref{1.5}), (\ref{1.6}) much more is needed; for example we must know how to compose momenta for multiparticle systems, what is the form of conservation laws, etc.  I will discuss these issues in the following sections below.

The  algebras (\ref{1.5}), (\ref{1.6}) has  been derived from the limit of the algebra of symmetries of three dimensional gravity, which, as it is well known, has some remarkable features, namely it is a topological field theory with no dynamical degrees of freedom. The question arises as to if it is possible to repeat this analysis in the most interesting, four dimensional case. One can expect that this latter case would be much more complex: to go to the appropriate limit reminding the Special Relativistic setting one should first switch off the dynamical degrees of freedom of gravity. The good news is that in the limit, in which the gravitational constant goes to zero, four dimensional gravity becomes a topological field theory again, reminding the three-dimensional situation. However, I must admit that it is not known if there exists a limit of four dimensional quantum gravity, which results in DSR theory. There are, however, some circumstantial evidences in favor of such a claim.

In the four-dimensional case the excitations of ground state\footnote{We restrict our attention to the ground state, because we are interested only in the limit in which all local degrees of freedom of quantum gravity are switched off. After all our goal is to formulate a theory which is to replace Special Relativity!} of a quantum gravity \index{quantum gravity} theory are
conjectured  to transform under representations of the quantum deformed
de Sitter algebra $SO_q(3,2)$,
with
$z=\ln q$ behaving in the limit of small $\Lambda \kappa^{-2}$ as,
$z  \approx  \Lambda \kappa^{-2}$  
\cite{linking}, \cite{baez-deform}, \cite{qdef}, \cite{artem}\footnote{From now on I put $\hbar$ equal 1.}. Then (see \cite{Amelino-Camelia:2003xp} for more details) one again takes the limit, which this time is much more involved, since one must not only rescale variables, as it was done above, but also to renormalize them (see also \cite{Lukierski:1991pn}), in order to get finite result. It turns out that now we have to do with one parameter family of contractions, labelled by real, positive parameter $r$: for $0<r<1$ as a result of contraction one obtains the standard Poincar\'e algebra, for $r>1$ the contraction does not exists and only for a single value $r=1$ the contraction gives the desired four dimensional $\kappa$-Poincar\'e algebra. It remains therefore an open problem whether and how quantum gravity singles out the value for $r$ and is this value 1?

We see therefore that it is possible to obtain the DSR1 algebra by contracting the algebras of symmetries of quantum gravity, in dimensions 3 and 4. This strongly suggests that indeed this algebra would be an algebra of symmetries of particle kinematics taking part in the flat space. It is interesting therefore that, in some cases at least, there are traces of quantum gravity in this algebra. I must stress, however that it remains to prove rigorously that the algebra $SO_q(3,2)$ indeed plays the conjectured role in quantum gravity. 
\newline

And now  something completely different. In Special Relativity the Poincar\'e algebra plays dual role: it is an algebra of symmetries of space--time and at the same time it labels momenta and spin of a particle. Deformed Poincar\'e algebra should also play such a dual role, so now let us investigate the algebras of charges carried by point particles coupled to quantum gravity. 
As I will show in section 3 below, in the DSR framework it turns out that the four momentum of a particle is not a point in the flat Minkowski space, as in Special Relativity, but instead, the manifold of momenta is a curved manifold of constant curvature, $\kappa^{-2}$ \cite{Kowalski-Glikman:2002ft}, \cite{Kowalski-Glikman:2003we}. But then, by the same token, positions, which are identified with ``translations'' of momenta, cannot commute, so that the space-time of DSR should necessarily be a non-commutative manifold, called $\kappa$-Minkowski space-time \cite{kappaM}, \cite{Kowalski-Glikman:2002jr}. Let us see therefore, how this picture emerges from quantum gravity, this time coupled to point particles, and without cosmological constant.

In what follows I will review the results obtained in \cite{Freidel:2003sp}.  Let us start with the case of three-dimensional quantum gravity now coupled to a point particle. Then it is well known (see the detailed and clear exposition in \cite{Matschull:1997du} and references therein) that
since in $3d$ gravity does not have any dynamical degrees of  freedom, the theory is fully characterized by Poincar\'e charges carried by the particle. In other words the theory reduces to a theory of the phase space of the particle, which is different from the phase space of free particles, as a result of the modifications induced by topological degrees of freedom of gravity. This phase space is characterized by the following properties \cite{Matschull:1997du}

\begin{itemize}
\item The coordinates of the particle (understood as variables on the phase space, which are canonically conjugated to momenta) do not commute and instead
\begin{equation}\label{1.7}
    [x_0, x_i] = -\frac{1}{\kappa}\, x_i, \quad [x_i, x_j] = 0.
\end{equation}
(The bracket above is either the Poisson bracket or the commutator.) Such a non-commutative space-time is called $\kappa$-Minkowski.
\item The space of (three-) momenta is not the flat ${R}^3$ manifold, but  the maximally symmetric space of constant curvature $-\kappa$ (anti de Sitter space of  momenta).
\item Last but not least it has been shown in numerous works on $3d$ quantum gravity that the full Hopf $\kappa$-Poincar\'e algebra with all the quantum group structures plays the role (see e.g., \cite{Freidel:2004vi} and references therein.)
\end{itemize}

But as I will show in Chapter 3, these are exactly the properties of phase space of a particle in DSR (in the case of both 3 and 4 dimensional space-time.) Note in passing an interesting duality between curvature and non-commutativity\footnote{See the insightful discussion in \cite{majidbook}, in which Shahn Majid argues that this duality indicates a deep relation between non-commutativity and quantization of gravity.}
$$
\begin{array}{c}
  \mbox{Curvature of momentum space} \\
  \Updownarrow \\
  \mbox{Non-commutativity of position space} \\
\end{array}
 $$
 
As I will show below this duality can be understood as a consequence of the co-product structure of quantum Poincar\'e algebra.

Thus we see again that kinematics of particles in three dimensions is described by the DSR-like structure with observer independent scale. The question arises as to if something similar can happen in four space-time dimensions. I have only circumstantial evidences in favor of such claim, and the argument goes as follows \cite{Freidel:2003sp}.
\newline

The main idea is to construct an experimental  situation that
forces a dimensional reduction from the four dimensional to the $2+1$ dimensional theory. It
is interesting that this can be done in quantum theory, using the
uncertainty principle as an essential element of the argument.
Let us consider a free 
elementary particle in $3+1$ dimensions, whose mass is less
than $G^{-1}= \kappa$.  The motion of the particle will be
linear, at least in some classes of coordinates systems, not accelerating with respect to the natural inertial coordinates at infinity.  Let us consider the particle as described by an inertial
observer who travels perpendicular to the plane of its
motion, which I will call the $z$ direction.  From the point of
view of that observer, the particle is in an eigenstate of 
longitudinal momentum, $\hat{P}^{total}_z$, with some eigenvalue
$P_{z}$. Since the particle is in an eigenstate of
$\hat{P}^{total}_z$ its wavefunction  will be
uniform in $z$, with
 wavelength $L$ where (note that I assume here that $L$ is so large that I can trust the standard uncertainty relation; besides  this uncertainty relation is not being modified in some formulations of DSR)
\begin{equation}
L= {1 \over P_{z}^{total} }
\end{equation}

At the same time, we assume that the  uncertainties in the
transverse positions are bounded a scale $r$,
such that $ r \ll 2L $.
Then the wavefunction for the the particle has support on a
narrow cylinder of radius $r$ which extend
uniformly in the $z$ direction.
Finally, we assume that the state of the  gravitational field is
semiclassical, so that to a good approximation, within $\cal C$
the semiclassical Einstein equations hold.
\begin{equation}
G_{ab}= 8\pi G <\hat{T}_{ab} >
\label{semi}
\end{equation}

Note that we do not have to assume that the  semiclassical
approximation holds for all states. We assume something much
weaker, which is that there are subspaces of states in which it
holds. This assumption is, in a sense, analogous to the assumption above that we are interested only in the analysis of ground state of quantum gravity.

Since the wavefunction is uniform in $z$, this implies that
the gravitational field seen by our observer will have a spacelike Killing
field $k^a= (\partial /\partial z)^a$.

Thus, if there are no forces other than the gravitational field, the
 particle described semiclassically by
(\ref{semi}) must be described by an equivalent $2+1$ dimensional
problem in which the gravitational field is dimensionally reduced
along the $z$ direction so that the particle, which
is the source of the gravitational field, is replaced by a
punctures.

The dimensional reduction is governed by a length $d$, which is
the extent in $z$ that the system extends. We cannot take $d<L$
without violating the uncertainty principle. It is then convenient
to take $d=L$.  Further, since the system consists of the
particle, with no intrinsic extent, there is no other
scale associated with their extent in the $z$ direction. We can
then identify $z=0$ and $z=L$ to make an equivalent toroidal
system, and then dimensionally reduce along $z$. The relationship
between the four dimensional Newton's constant $G^{4}$ and the
three dimensional Newton's constant $G^{3}=G$ is given by
\begin{equation}
G^{3} = {G^{4} \over L} = {G^{4} P^{tot}_z \over \hbar}
\end{equation}

Thus, in the analogous $3$ dimensional system, which is equivalent
to the original system as seen
from the point of view of the boosted observer, the Newton's
constant depends on the longitudinal momentum.

Of course, in general there
will be an additional scalar field, corresponding to the
dynamical degrees of freedom of the gravitational field. We will
for the moment assume that these are unexcited, but exciting them
will not affect the analysis so long as the gravitational
excitations are invariant also under the Killing field and are of
compact support.

Now we note that, if there are no other particles or excited
degrees of freedom, the energy of the system  can
to a good approximation be described by the hamiltonian $H$ of the
two dimensional dimensionally reduced system. This is described
by a boundary integral, which may be taken over any circle that
encloses the particle.
But it is well known that in $3d$ gravity $H$ is bounded
from above. This may seem strange, but it is easy
to see that it has a natural four dimensional interpretation.

The bound is given by
\begin{equation}
M < {1 \over 4 G^{3} } = {L \over 4 G^{4} }
\end{equation}
where $M$ is the value of the ADM hamiltonian, $H$. But this just
implies that
\begin{equation}
L > 4G^{4}M = 2R_{Sch} \label{sch}
\end{equation}
i.e. this has to be true, otherwise the dynamics of the
gravitational field in $3+1$ dimensions would have collapsed the
system to a black hole!  Thus, we see that the total bound from
above of the energy in $2+1$ dimensions is necessary so that one
cannot violate the condition in $3+1$ dimensions that a system be
larger than its Schwarzschild radius.

Note that we also must have
\begin{equation}
M > P^{tot}_z ={
\hbar \over L}
\end{equation}
Together with (\ref{sch}) this implies $L>
l_{Planck}$, which is of course necessary if the semiclassical
argument we are giving is to hold.

Now, we have put no restriction on any components of  momentum or
position in the transverse directions.  So the system still has
symmetries in the transverse directions.  Furthermore, the argument
extends to any number of particles, so long as their relative
momenta are coplanar. Thus, we learn the following.

Let ${\cal H}^{QG}$ be the full Hilbert space of the quantum
theory of gravity, coupled to some appropriate matter fields, with
$\Lambda=0$. Let us consider a subspace of states ${\cal
H}^{weak}$ which are relevant in the low energy limit in which all
energies are small in Planck units.  We expect that this will have
a symmetry algebra which is related to the Poincar\'e algebra
${\cal P}^{4}$ in $4$ dimensions, by some possible small
deformations parameterized by $G^{4}$ and $\hbar$. Let us call
this low energy symmetry group ${\cal P}^{4}_{G}$.

Let us now   consider the subspace of ${\cal H}^{weak}$ which is
described by the system we have just constructed . It contains the
particle, and is an eigenstate of $\hat{P}^{tot}_z$ with large
$P^{tot}_z$ and vanishing  longitudinal momentum.
Let us call this subspace of Hilbert space
${\cal H}_{P_z}$.

The conditions that define this subspace break the  generators of
the (possibly modified) Poincar\'e algebra that involve the $z$
direction. But they leave unbroken the symmetry in the $2+1$
dimensional transverse space. Thus, a subgroup of ${\cal
P}^{3+1}_{G}$ acts on this space, which we will call ${\cal
P}^{2+1}_{G} \subset {\cal P}^{3+1}_{G}$.

We have argued that the physics in ${\cal H}_{P_z}$ is to good
approximation described by an analogue system in of a particle
in $2+1$ gravity. However, we know from the results mentioned above that the symmetry algebra acting there is not 
the ordinary $3$ dimensional Poincar\'e algebra, but  the
$\kappa$-Poincar\'e algebra in $3$ dimensions, with
\begin{equation}
\kappa^{-1}  = {4 G^{4} P^{tot}_z \over \hbar}
\end{equation}

Now we can note the following. Whatever ${\cal P}^{4}_{G}$ is,
it must have the following properties:

\begin{itemize}
\item{} It depends on $G^{4}$ and $\hbar$, so that it's  action on
{\it each} subspace ${\cal H}_{P_z}$, for each choice of $P_z$, is
the $\kappa$ deformed $3d$ Poincar\'e algebra, with $\kappa$ as
above.

\item{} It does not satisfy the rule that momenta and energy add, on
all states in $\cal H$, since they are not satisfied in these
subspaces.

\item{} Therefore, whatever $ {\cal P}^{4}_{G}$ is, it is not the
classical Poincar\'e group.
\end{itemize}

Thus the theory of particle kinematics at ultra high energies is
not Special Relativity, and the arguments presented above suggest
that it might be Doubly Special Relativity. \index{Doubly Special Relativity} So it is good time now to start discussing the structures of this theory.

\section{Doubly Special Relativity and the $\kappa$-Poincar\'e algebra}
\def\bbbone{{\mathchoice {\rm 1\mskip-4mu l} {\rm 1\mskip-4mu l}
{\rm 1\mskip-4.5mu l} {\rm 1\mskip-5mu l}}}
\def\vP{\vec{P}\, {}^2}

Soon after pioneering papers of Amelino-Camelia \cite{Amelino-Camelia:2000ge}, \cite{Amelino-Camelia:2000mn} it was realized in
\cite{Kowalski-Glikman:2001gp} and \cite{Bruno:2001mw} that the 
$\kappa$-Poincar\'e algebra \index{$\kappa$-Poincar\'e algebra} \cite{Lukierski:1991pn}, \cite{Lukierski:1991ff}, \cite{kappaM} is a perfect
mathematical setting to describe one particle kinematics in DSR. Let us recall from the preceding section that in particular,
in the  bicrossproduct basis the brackets of rotations $M_i$,
boosts $N_i$, and the components of momenta $P_\mu$ read\footnote{From now on I will be discussing the four-dimensional case only. However, the reader can easy convince her(him)self that what will be said here applies with minor and obvious modifications in other dimensions as well. Notice that now I use the QFT convention of adding the ``$i$'' on the right-hand-side of the algebra.}
$$
[M_i, M_j] = i\, \epsilon_{ijk} M_k, \quad [M_i, N_j] = i\, \epsilon_{ijk} N_k,
$$
\begin{equation}\label{1}
  [N_i, N_j] = -i\, \epsilon_{ijk} M_k,
\end{equation}
\begin{equation}\label{2}
  [M_i, P_j] = i\, \epsilon_{ijk} P_k, \quad [M_i, P_0] =0,
\end{equation}
\begin{equation}\label{3}
   \left[N_{i}, P_{j}\right] = i\,  \delta_{ij}
 \left( {\kappa\over 2} \left(
 1 -e^{-2{P_{0}/ \kappa}}
\right) + {1\over 2\kappa} \vec{P}\,{}^{ 2}\, \right) - i\, {1\over \kappa}
P_{i}P_{j}
\end{equation}
\begin{equation}\label{4}
\left[N_{i}, P_{0}\right] = i\, P_i
\end{equation}

It is important to note that the algebra  of $M_i$ $N_i$ is
just the standard Lorentz algebra, so one of the first conclusions is that the
Lorentz sector of $\kappa$-Poincar\'e algebra is not deformed. Therefore in DSR
theories, in accordance with the first postulate above, the Lorentz symmetry
{\em is not broken} but merely nonlinearly realized in its action on momenta.
This simple fact has lead some authors (see e.g., \cite{lunoDSR},
\cite{Ahluwalia:2002wf}) to the
 claim that DSR is nothing but the standard Special Relativity in non-linear
disguise. As we will see this view is clearly wrong, simply because the algebra
(\ref{1})--({\ref{4}) describes only half of the phase space of the particle,
and the full phase space algebra cannot be reduced to the one of Special
Relativity.

As one can easily check, the Casimir of the $\kappa$-Poincar\'e algebra reads
\begin{equation}\label{5}
\kappa^2\,  \cosh\, \frac{P_0}{\kappa} - \frac{\vec{P}{}^2}2\, e^{P_0/\kappa} = M^2.
\end{equation}

{\bf Exercise 5}. Check that (\ref{5}) is indeed the Casimir of the algebra (\ref{1})--(\ref{4}) i.e., its commutators with all the generators of $\kappa$-Poincar\'e algebra vanish. Is it the only possible Casimir of this algebra? Compute the velocity $v = \partial P_0/\partial |\vec P|$. How the behavior of this velocity depends on the sign of $\kappa$?\newline 

It follows from (\ref{5}) that the value of three-momentum $|\vec{P}|=\kappa$
corresponds to  infinite energy $P_0=\infty$. One can check easily (see Exercise 4 above) that in
this particular realization of DSR $\kappa$ is indeed observer independent
\cite{Kowalski-Glikman:2001gp}, \cite{Bruno:2001mw} (i.e., if a particle has momentum 
$|\vec{P}|=\kappa$ for some observer, it has the same momentum for all, Lorentz related, 
observers.) One also sees that the
speed of massless particles, naively defined as derivative of energy over
momentum, increases monotonically with momentum and diverges for the maximal
momentum $|\vec{P}|=\kappa$, if $\kappa$ is positive. As I mentioned already in the DSR terminology, the theory based on the algebra (\ref{1})--(\ref{4}) with Casimir (\ref{5}) is sometimes called DSR1.\index{DSR1}

One should note at this point that the bicrossproduct algebra above is not the
only possible realization of DSR. For example, in \cite{Magueijo:2001cr},
\cite{Magueijo:2002am} Magueijo and Smolin proposed and
carefully analyzed another DSR proposal, called sometimes DSR2.\index{DSR2} 
In DSR2 the Lorentz algebra is still not deformed and there are no deformations in the brackets of rotations and momenta. The boosts-- momenta generators have now the form
\begin{equation}\label{ms1}
   \left[N_{i}, p_{j}\right] =  i\left( \delta_{ij}p_0 -
  {1\over \kappa} p_{i}p_{j} \right),
\end{equation}
and
\begin{equation}\label{ms2}
  \left[N_{i},p_{0}\right] = i\, \left( 1 - {p_0\over \kappa}\right)\,p_{i}.
\end{equation}
It is easy to check that the Casimir for this algebra has the form
\begin{equation}\label{ms3}
 M^2 = \frac{p_{0}^2 - \vec{p}{}^2}{\left(1- \frac{p_0}\kappa\right)^2}.
\end{equation}

{\bf Exercise 6}. Check that (\ref{ms3}) is indeed the Casimir of the DSR2 algebra (\ref{ms1}), (\ref{ms2}).  Compute the velocity $v = \partial P_0/\partial |\vec P|$. Find relations between DSR1 and DSR2 momentum variables (the answer can be found in \cite{Kowalski-Glikman:2002we},
\cite{Kowalski-Glikman:2002jr}.)\newline

Moreover there is a basis of DSR, closely related to
the famous Snyder theory \cite{snyder}, in which the energy-momentum space
algebra is purely classical (it was first found in \cite{Kosinski:1994br} and further analyzed in \cite{Kowalski-Glikman:2002we},
\cite{Kowalski-Glikman:2002jr}.)\newline

{\bf Exercise 7}. Find explicit transformation from DSR1 to the classical basis, in which all the brackets are identical to those of the standard Poincar\'e algebra. (See \cite{Kowalski-Glikman:2002we},
\cite{Kowalski-Glikman:2002jr}, where the relation of the DSR algebra in classical basis and  Snyder's theory is analyzed in details.)

\subsection{Space-time of DSR}

The formulation of DSR in the energy-momentum space is clearly incomplete, as it
lacks any description of the structure of space-time.  DSR
has been formulated in a somehow  unusual way:  one started with the energy--momentum space and only then the problem of construction of space-time had been considered. Usually we do the opposite, for example in the standard formulation of Special Relativity one starts with clear operational definition of space-time notions (distance, time interval) and only then the energy-momentum space and phase space is being constructed.\newline

{\bf Exercise 7}. (Difficult\footnote{By this I mean that I do not quite know how to solve it (as a matter of fact I believe nobody does)!}.) Formulate Special Relativity in the operational way, taking as a starting point the space of energy and momenta.\newline

There are in principle many ways how the phase space can be constructed. For
example in \cite{Kimberly:2003hp} one constructs the position space along the
same lines as the energy-momentum space has been constructed in
\cite{Magueijo:2001cr}, \cite{Magueijo:2002am}. Here, following \cite{Kowalski-Glikman:2002jr}, I take another route. As I have been stressing in the preceding section,
 one of the distinctive features of the $\kappa$-Poincar\'e algebra\index{$\kappa$-Poincar\'e algebra}
is that it possesses additional structures that make it a Hopf algebra. Namely
one can construct the so called co-products for the rotation, boosts, and momentum
generators, which, in turn, can be used to provide a procedure to construct the
phase space in a unique way.

The co-product \index{co-product} is the mapping from the algebra ${\cal A}$ to the tensor product ${\cal A}\otimes {\cal A}$
satisfying some requirements that make it in a sense dual to algebra multiplication (see \cite{majidbook} for details), which essentially provides a rule how the algebra acts on products (of functions, and, in physical applications, on multiparticle states.)
For the bicrossproduct $\kappa$-Poincar\'e algebra (\ref{1})--(\ref{4}) the co-products read
\begin{equation}\label{6}
 \Delta (P_0) = \bbbone\otimes P_0 + P_0 \otimes \bbbone
\end{equation}
\begin{equation}\label{7}
\Delta (P_k) = P_k\otimes {\rm e}^{-P_0/\kappa} + \bbbone \otimes P_k
\end{equation}
\begin{equation}\label{8}
  \Delta
(M_i) = M_i\otimes \bbbone + \bbbone \otimes M_i
\end{equation}
\begin{equation}\label{9}
  \Delta
(N_i) = \bbbone\otimes N_i + N_i \otimes {\rm e}^{-P_0/\kappa} -
\frac{1}{\kappa}\epsilon_{ijk}\, M_j \otimes P_k
\end{equation}

In order to construct the one-particle phase space we must first introduce
objects that are dual to $M_i$, $N_i$, and $P_\mu$. These are the matrix
$\Lambda^{\mu\nu}$ and the vector $X^\mu$. Let us briefly interpret their
physical meaning. $X^\mu$ are to be dual to momenta $P_\mu$, which clearly
indicates that they should be interpreted as translation of momenta, in other words the positions. The duality between $\Lambda^{\mu\nu}$
and $M_{\mu\nu} = (M_i, N_i)$ is a bit more tricky. However if one interprets
$M_{\mu\nu}$ in analogy to the interpretation of momenta, i.e., as Lorentz
charge carried by the particle, that is its angular momentum, then the dual
object $\Lambda^{\mu\nu}$ has clear interpretation of Lorentz transformation.
Thus we have the structure of the form $G \times {\cal MP}$, where $G$ is the
Poincar\'e group acting on the space of Poincar\'e charges of the particle
${\cal MP}$. We see therefore that we can make use of the powerful
mathematical theory of Lie-Poisson groups and co-adjoint orbits (see, for
example, \cite{kirillov:1976}, \cite{alekseev:1994}) and their quantum deformations.

Following \cite{kosinski1995} and \cite{luno} we assume the following form of
the co-product on the group
\begin{equation}\label{10}
 \Delta(X^{\mu})=\Lambda^{\mu}{}_{\nu}\otimes
X^{\nu}+X^{\mu}\otimes \bbbone
\end{equation}
and
\begin{equation}\label{11}
\Delta(\Lambda^{\mu}{}_{\nu})=\Lambda^{\mu}{}_{\rho}\otimes
\Lambda^{\rho}{}_{\nu}
\end{equation}
The next step is to define the pairing between elements of the algebra and of the group in a canonical way that establish 
the duality between these two structures.
\begin{equation}\label{12}
 <P_{\mu},X^{\nu}>= i\delta_{\mu}^{\nu}
\end{equation}
\begin{equation}\label{13}
  <M^{\alpha\beta},\Lambda^{\mu}{}_{\nu}>= i
\left(g^{\alpha\mu}\delta^{\beta}_{\nu}-g^{\beta\mu}\delta^{\alpha}_{\nu}\right)
\end{equation}
\begin{equation}\label{14}
  <\Lambda^{\mu}{}_{\nu},1>= \delta^{\mu}_{\nu}
\end{equation}
In (\ref{13}) $g^{\alpha\mu}$ is the Minkowski space-time metric. This pairing
must be consistent with the co-product structure in the following sense
\begin{equation}\label{16a}
  <A, XY> = <A_{(1)}, X><A_{(2)}, Y>,
\end{equation}
 \begin{equation}\label{16b}
 <AB,X> =<A, X_{(1)}><B, X_{(2)}>,
\end{equation}

The rules (\ref{12})--(\ref{16b}) make it possible to construct the commutator
algebra of the phase space. To this end one makes use of the Heisenberg double\index{Heisenberg double}
procedure \cite{alekseev:1994}, \cite{luno}, that defines the brackets in terms
of the pairings as follows (no summation over repeated indices here!)
\begin{equation}\label{15}
 \left[X^{\mu},P_{\nu}\right]= P_{\nu(1)}\left< X^{\mu}_{(1)},P_{\nu(2)}\right> X^{\mu}_{(2)}
-P_{\nu}X^{\mu},
\end{equation}
\begin{equation}\label{16}
  \left[X^\mu,M^\rho{}_\sigma\right]= M_{(1)}{}^\rho{}_\sigma \left< X_{(1)}{}^\mu,M_{(2)}{}^\rho{}_\sigma \right> X_{(2)}{}^\mu - M^\rho{}_\sigma X^\mu,
\end{equation}
and analogously for $\Lambda^{\mu}{}_{\nu}$ commutators, where on the right
hand side we make use of the standard (``Sweedler'') notation for co-product
$$
\Delta {\cal T} = \sum {\cal T}_{(1)} \otimes {\cal T}_{(2)}.
$$

As an example let us perform these steps in the case of the bicrossproduct $\kappa$-Poincar\'e algebra of DSR1. It follows
from (\ref{7}), and (\ref{12}), and (\ref{16b}) that
$$
<P_i, X_0 X_j> = -\frac1\kappa\, \delta_{ij}, \quad <P_i,  X_jX_0>=0,
$$
from which one gets
\begin{equation}\label{17a}
[X_0, X_i] \equiv X_0 X_i - X_i X_0= -\frac{i}\kappa\, X_i.
\end{equation}
Similarly, using (\ref{15}) we get the standard relations
\begin{equation}\label{17b}
[P_0, X_0] = -i, \quad [P_i, X_j] = i \, \delta_{ij}.
\end{equation}
It turns out that the phase space algebra contains one more non-vanishing
commutator (which can be, of course, also obtained from Jacobi identity),
namely
\begin{equation}\label{17c}
 [P_i, X_0] = -\frac{i}\kappa\, P_i.
\end{equation}

Thus we have constructed the phase space of the bicrossproduct $\kappa$-Poincar\'e algebra of DSR1. Let us
stress that this construction relies heavily on the form of co-product.
However, as it will turn out below, some of the commutators are
sensitive to the particular form of the DSR, while the others are not. In
particular we will see that the non-commutativity of positions (\ref{17a}) is
to large extend universal for a whole class of DSR theories. The
non-commutative space-time with such Lie-like type of non-commutativity is
called $\kappa$-Minkowski space-time. \index{$\kappa$-Minkowski space-time}
\newline

{\bf Exercise 8}. Using Jacobi identity derive  the brackets of boosts and positions, assuming that they form a Lie algebra. Which algebra is it? (The answer can be found below.)

\subsection{From DSR theory to DSR theories}

The introduction of invariant momentum (or mass) scale $\kappa$ has immediate
consequences.  The most important is that there is nothing sacred about the
bicrossproduct DSR presented above, as one can simply use
$\kappa$ to define new energy and momentum (new basis of DSR) as analytic functions of the old
ones, to wit
\begin{equation}\label{18}
{ \cal P}_i = {\cal F}_i(P_i, P_0; \kappa), \quad { \cal P}_0 = {\cal F}_0(P_i,
P_0; \kappa),
\end{equation}
the only restrictions being that the equations in (\ref{18})  transform covariantly under rotations and that in the $\kappa\rightarrow\infty$ limit ${ \cal P}_\mu = P_\mu$, because we insist on the right low energy limit in all the bases.
Observe that such a change of energy and momentum  is not possible in a theory without
any mass scale, like special relativity and Newtonian mechanics,  in which the
energy momentum spaces are linear, and the mass shell conditions are expressed by
quadratic form. 

Then  a natural
question arises: which momenta are the ``right'' ones? The hope is that the
theory of quantum gravity or some other fundamental theory, from which DSR is
descending will tell what is the correct physical choice. One can also contemplate the
possibility that in the final, complete formulation of DSR one will have to do
with some kind of ``energy-momentum general covariance'', i.e., that physical
observables do not depend on a particular realization of eq.~(\ref{18}), like
observables in general relativity do not depend on coordinate system. Then a
natural question arises: is it possible to understand transformations
(\ref{18}) as coordinate transformations on some (energy-momentum) space?

Surprisingly enough the answer to this question is in the positive: indeed the
transformations between DSR theories, described by (\ref{18}) are nothing but
coordinate transformation of the constant curvature manifold, on which momenta
live. To reach this conclusion one observes first
\cite{Kowalski-Glikman:2002we}, \cite{Kowalski-Glikman:2002jr} that it follows
from the Heisenberg double construction that both the $\kappa$-Minkowski
commutator (\ref{17a}) and the commutators between Lorentz charges $M_{\mu\nu}$
and positions $X_\mu$ are left invariant by the transformations (\ref{18}).
This follows from the fact that the transformations (\ref{18}) a severely
constrained by assumed rotational invariance and the fact that in the $\kappa\rightarrow\infty$ limit the new energies and momenta must must be the same as in the standard Special Relativity. Since the bicrossproduct DSR variables satisfy this requirement it follows that the new variables cannot differ from the DSR1 ones in the $\kappa^0$ order. Therefore, in the leading order, they must be of the form
\begin{equation}\label{19}
{ \cal P}_i \approx P_i + \alpha\, \frac1\kappa\, P_i P_0 +
O\left(\frac1{\kappa^2}\right), \quad { \cal P}_0 = P_0 + \beta\,
\frac1\kappa\, P_0^2 + O\left(\frac1{\kappa^2}\right)
\end{equation}
where $\alpha$ and $\beta$ are numerical parameters. It turns out that in
computing  the brackets of positions $X$ and the ones of positions with boosts Heisenberg double procedure picks up only
the first terms in this expansion, and thus the form of the commutators remains
unchanged. Of course, the position-momenta commutators are changed by the transformations
(\ref{18}), (\ref{19}).\newline

{\bf Exercise 9}. Using expansion (\ref{19}) derive the brackets of positions and four-momenta ${ \cal P}_\mu$. It would help to notice that co-product is a homomorphism and thus $\Delta(ab)= \Delta(a)\Delta(b)$.\newline

Next it was realized in \cite{Kowalski-Glikman:2002ft},
\cite{Kowalski-Glikman:2003we} that the algebra of positions and Lorentz
charges is nothing but  de Sitter $SO(4,1)$ algebra. \index{de Sitter algebra} The positions and Lorentz transformations are, in turn, nothing but the transformations of the manifold, whose points are energy and momenta (energy-momentum manifold.) On this manifold positions are generators of translational symmetry, while boosts and rotations generate Lorentz transformations. Thus the energy--momentu manifold is a four-dimensional manifold with ten-parameter group of symmetries and thus it must be a maximally symmetric space of constant curvature. It follows from the well known theorem of differential geometry that such a manifold must be locally diffeomorphic to one of the three spaces of constant curvature, and since the group of symmetries is $SO(4,1)$, this manifold must be de Sitter space\footnote{It turns out that all other spaces of constant curvature are also possible, if one generalizes somehow the definition of $\kappa$-Poincar\'e algebra, i.e., the phase space associated with $\kappa$-Poincar\'e algebra can have positive, zero, and negative curvature (see \cite{Blaut:2003wg} for details.)}. Then it follows that the algebra of positions and Lorentz transformations is just an algebra of symmetries of de Sitter space, and therefore it is, of course, independent of a coordinate system we use to describe this space.

De Sitter space of momenta \index{de Sitter space of momenta} can be constructed as a four dimensional  surface of constant
curvature $\kappa$ in the five dimensional Minkowski space with coordinates
$\eta_A$, $A=0,\ldots,4$, to wit
\begin{equation}\label{20}
-\eta_0^2 + \eta_1^2 + \cdots + \eta_4^2 = \kappa^2.
\end{equation}
The $SO(4,1)$ generators can be decomposed into positions $X_\mu$ and Lorentz charges $M_{\mu\nu}$, which act on $\eta_A$ variables as
follows
\begin{equation}\label{21}
  [X_0,\eta_4] = \frac{i}\kappa\, \eta_0, \quad [X_0,\eta_0] = \frac{i}\kappa\, \eta_4, \quad [X_0,\eta_i] = 0,
\end{equation}
\begin{equation}\label{22}
  [X_i, \eta_4] = [X_i, \eta_0] =\frac{i}\kappa\, \eta_i, \quad [X_i, \eta_j] = \frac{i}\kappa\,
\delta_{ij}(\eta_0 - \eta_4),
\end{equation}
and
\begin{equation}\label{23}
  [M_i, \eta_j] = i\epsilon_{ijk}\eta_k, \quad [N_i, \eta_j] = i\, \delta_{ij}\, \eta_0, \quad [N_i, \eta_0] = i\,  \eta_i,
\end{equation}
It should be noted that there is another decomposition of $SO(4,1)$ generators
\cite{Kowalski-Glikman:2002ft}, \cite{Kowalski-Glikman:2003we}, in which the
resulting algebra is exactly the one considered by Snyder \cite{snyder}.

On the space (\ref{20}) one can built various co-ordinate systems, each related
to some DSR theory. In particular, one recovers the bicrossproduct DSR1 \index{DSR1} with
the following coordinates (which are, accidentally, the standard ``cosmological'' coordinates
on de Sitter space)
\begin{eqnarray}
{\eta_0} &=& -\kappa\, \sinh \frac{P_0}\kappa - \frac{\vec{P}\,{}^2}{2\kappa}\,
e^{  \frac{P_0}\kappa} \nonumber\\
\eta_i &=&   - P_i \, e^{  \frac{P_0}\kappa} \nonumber\\
{\eta_4} &=&  \kappa\, \cosh \frac{P_0}\kappa  - \frac{\vec{P}\,{}^2}{2\kappa}
\, e^{  \frac{P_0}\kappa}.   \label{24}
\end{eqnarray}
Using (\ref{24}), (\ref{22}), and the Leibnitz rule, one easily recovers the
commutators (\ref{1})--(\ref{4}).\newline

{\bf Exercise 10}. Check this explicitly.\newline

Other coordinates systems, are possible, of course.\newline

{\bf Exercise 11}. Find the coordinates on de Sitter space of momenta, corresponding to DSR2.\newline 

 In particular one can
choose the ``standard basis'' in which
\begin{equation}\label{25}
 {\cal P}_\mu = \eta_\mu/\eta_4.
\end{equation}
Note that in this basis (or classical DSR) the commutators of all Poincar\'e
charges, ${\cal P}_\mu$ and $M_{\mu\nu}$ are purely classical. However, the
positions brackets, as well as the momenta/positions
cross-relations are still non-trivial.
\newline

{\bf Exercise 12}. Compute the bracket of positions with energy and momenta in the classical basis.\newline

This means that in the classical bases of DSR  the (observer-independent) scale $\kappa$
disappears completely from the Lorentz sector, but is still present in the
translational one. Thus such a theory fully deserves the name DSR. 

De Sitter space setting reveals the geometrical structure of DSR theories. As
we saw the energy momentum  space of DSR is a four dimensional manifold of
positive constant curvature, and the curvature radius equals the scale
$\kappa$. The Lorentz charges and positions are identified with the set of ten
tangent vectors to the de Sitter energy-momentum space, and as an immediate
consequence of this their algebra is independent of any particular coordinate
system on this space.  However the latter seems to be, at least naively,
physically relevant. Each such coordinate system defines for us (up to the
redundancy discussed in \cite{Kowalski-Glikman:2003we}) the physical  energy and momentum. In one-particle sector the particular choice may not be
relevant, but it seems that it would be of central importance for the proper
understanding of many particles phase spaces, in particular in analysis of the
phenomenologically important issue of particles scattering and conservation
laws.

Having obtained the one-particle phase space of DSR, it is natural to proceed
with construction of the field theory. Here two approaches are possible. One
can try to construct field theory on the non-commutative $\kappa$-Minkowski
space-time. Attempts to construct such a theory has been reported, for example,
in 
\cite{Kosinski:2003xx} and references therein, as well as in \cite{Amelino-Camelia:2001fd},
\cite{Amelino-Camelia:2002mu}. This line of research is, however, far from
being able to give any definite results, though some partial results, like an
interesting, nontrivial vertex structure reported in
\cite{Amelino-Camelia:2001fd}, \cite{Amelino-Camelia:2002mu} may shed some
light on physics of the scattering processes. The major obstacle seems to be
lack of the understanding of functional analysis on the spaces with Lie-type of
non-commutativity, which is most likely a deep and hard mathematical problem
(already the definition of appropriate differential and integral calculi is a
mater of discussion.) Therefore it seems simpler (and in fact more along the
line of the DSR proposal, where the energy momentum space is more fundamental
than the space-time structures) to try to built (quantum) field theory in
energy-momentum space directly. This would amount to understand how to define
(quantum) fields on the curved energy-momentum space, but, in principle, for
spaces of constant curvature at least functional analysis is well understood.
It should be noted that such an idea has been contemplated for a long time, and
in fact it was one of the main motivations of \cite{snyder}. Field theories
with curved energy-momentum manifold has been intensively investigated by Kadyshevsky 
and others \cite{Kadyshevsky}, without any conclusive results, though.

\section{Physics with Doubly Special Relativity}

Till now I have been discussing formal aspects of Doubly Special Relativity in a particular formulation, in which quantum algebras and non-commutative space-time played the fundamental role. Now it is time to try to turn to more physical questions, related with possible experimental signatures of quantum gravity. In other contributions to this volume, the reader can find
much more detailed discussion of the ``quantum gravity phenomenology'', here I would like to concentrate on those physical aspects and problems that are directly related to a particular formulation of DSR in terms of $\kappa$-Poincar\'e algebras.

\subsection{Time-of-flight experiments and the issue of velocity in DSR}

One of the simplest experimental tests of quantum gravity phenomenology is the time-of-flight experiment. In this experiment which is to be performed in a near future with good accuracy by  GLAST satellite (see e.g.,
\cite{Amelino-Camelia:2002vw} and references therein) one measures the energy-dependence of velocity of light coming from a distant source. Naively, most DSR models predicts positive signal in such an experiment (for details see \cite{Amelino-Camelia:2003ex}.) Indeed, in DSR $\partial E(p)/\partial p$ does, with an exception of the classical bases, depend on energy, which suggest that velocity of massless particles may depend on the energy they carry. This is the case, for example, both in the bicrossproduct DSR1 and in the Magueijo-Smolin DSR2 model.

Of course, the velocity formula  should be derived from the first principles. In the careful analysis reported in \cite{Amelino-Camelia:2002tc} (based on the calculations presented some time ago in \cite{Amelino-Camelia:1999pm}) the authors construct the wave packet from plane waves moving on the $\kappa$-Minkowski space-time, and then calculate the group velocity of such a packet, which, they claim, turns-out to be exactly $v^{(g)} = \partial E(p)/\partial p$\footnote{Notice however that similar analysis presented in \cite{Tamaki:2002iz} resulted in different conclusion. I will discuss below the reason for this discrepancy.}. This result is puzzling in view of the phase space calculation of velocity, which I will present below. Therefore let us analyze this calculation in more details.

The authors of \cite{Amelino-Camelia:2002tc} consider the wave packet built of waves moving in non-commutative $\kappa$-Minkowski space-time, centered at  $(\omega_{0},\vec{k}_{0})$, to wit
\begin{equation}\label{w1}
   \Psi _{(\omega _{0},\vec{k}_{0})}(\vec{x},t)=\int e^{i\vec{k}{\cdot}
\vec{x}}e^{-i\omega t}d\mu 
\end{equation}
Here the plane waves have been ordered so that the time variable appear on the right, and $d\mu$ is an appropriate measure on the space of three-momenta, whose detailed form will be irrelevant to what follows., We assume that the plane waves in the integral satisfy appropriate field equations so that $\omega$ is a given function of $\vec{k}$ such that for the pair $(\vec{k}, \omega(\vec{k}))$ the Casimir vanishes identically. Let us assume that the integral in (\ref{w1}) has support on small neighborhood   $$\omega_{0}-\Delta \omega \leq \omega \leq \omega _{0}+\Delta \omega $$ $$
 \vec{k}_{0}-\Delta \vec{k}\leq \vec{k}\leq \vec{k}_{0}+\Delta \vec{k}$$
Factoring out the phases  $e^{i k_{0}x}$  to the left and  $e^{-i\omega _{0}t}$ to the right one gets
\begin{equation}\label{w2}
    \Psi^m_{(\omega _{0},\vec{k}_{0})}(\vec{x},t)=e^{i\vec{k}_{0}{\cdot}
\vec{x}}{\left[ \int e^{i\Delta \vec{k}{\cdot}\vec{x}}
e^{-i\Delta \omega t}d\mu \right] } e^{-i\omega _{0}t} 
\end{equation}
Now the integral in the middle carries the information about the group velocity of the wave packet. Indeed it follows that the group velocity equals (in deriving the expression above one should make use of the fact that in the limit $\Delta \omega, \Delta \vec{k} \rightarrow0$, the commutator $[e^{-i\Delta \vec{k}{\cdot}\vec{x}}, e^{i\Delta \omega t}]=0$)
\begin{equation}\label{w3}
    v^{(g)} = \lim_{\Delta \vec{k}\rightarrow0}\, \frac{ \Delta \omega}{ |\Delta \vec{k}|} = \frac{d  \omega}{d | \vec{k}|} = \frac{d  E}{d | \vec{P}|}.
\end{equation}

The expression (\ref{w2}) is, however, ambiguous because the middle, amplitude term does not commute with the exponents on the left and on the right as a result of the identity 
$$
e^{i\vec{k}{\cdot}
\vec{x}} e^{-i\omega t} = e^{-i\omega t}e^{i\,e^{-\omega/\kappa}\,\vec{k}{\cdot}
\vec{x}}
$$
Thus instead of (\ref{w2}) we can use
\begin{equation}\label{w4}
    \Psi^r_{(\omega _{0},\vec{k}_{0})}(\vec{x},t)=  e^{i\vec{k}_{0}{\cdot}
\vec{x}}\, e^{-i\omega _{0}t}\, \left[ \int e^{i\,e^{-\omega/\kappa}\, \Delta \vec{k}{\cdot}\vec{x}}
e^{-i\Delta \omega t}d\mu \right]
\end{equation}
or
\begin{equation}\label{w5}
    \Psi^l_{(\omega _{0},\vec{k}_{0})}(\vec{x},t)= \left[ \int e^{i\Delta \vec{k}{\cdot}\vec{x}}
e^{-i\Delta \omega t}d\mu \right]\, e^{i\vec{k}_{0}{\cdot}
\vec{x}}\, e^{-i\omega _{0}t} 
\end{equation}
where in the last expression, we neglected the $e^{-\Delta \omega/\kappa}$ term in the exponent (it goes to zero in the relevant limit.)

We see therefore that the group velocity depends on the ordering of the wave packet (\ref{w2}), (\ref{w4}), (\ref{w5}) and equals
\begin{equation}\label{w6}
    v^{(g)} = \left\{
    \begin{array}{ll}
      \frac{d  \omega}{d | \vec{k}|} & \quad\mbox{ in the cases m, l} \\
     \frac{d  \omega}{d | \vec{k}|} \, e^{-\omega/\kappa}  & \quad \mbox{ in the case r}  \\
    \end{array}\right.
\end{equation}
Using the fact that for massless particles $\omega$ and $\vec{k}$ are related by (see (\ref{5})
\begin{equation}\label{w7}
\kappa^2\,  \cosh\, \frac{\omega}{\kappa} - \frac{\vec{k}{}^2}2\, e^{\omega/\kappa} = \kappa^2.
\end{equation}
we find easily
\begin{equation}\label{w8}
    v^{(g)} = \left\{\begin{array}{ll}
      \frac{\kappa}{\kappa - |\vec{k}|} &\quad \mbox{ in the cases m, l} \\
     1  & \quad\mbox{ in the case r} \\
    \end{array}\right.
\end{equation}    
Thus we see that the ordering ambiguity in the derivation leads to the ambiguity in the prediction of DSR1 concerning one of the few effects that might be in principle observed. In particular, for one ordering we have velocity of massless particles growing with the energy, while for other we have constant speed of light, as in Special Relativity. The only way out, therefore, is to  compute the velocity in a different, though physically equally appealing framework.

To this aim let us try to compute the velocity starting from the phase space of DSR theories. This computation has been presented in \cite{Daszkiewicz:2003yr} (see also \cite{Kosinski:2002gu} and \cite{Mignemi:2003ab}.)

The idea is to start with the commutators (\ref{21})--(\ref{23}). Note first that since  
the for the variable $\eta_4$, $[M_i, \eta_4]=[N_i,\eta_4]=0$,  $\kappa\, \eta_4$ is a Casimir 
(cf.~(\ref{24})) and can be therefore naturally identified with the relativistic Hamiltonian ${\cal H}$ for \index{relativistic Hamiltonian in DSR}
free particle in any DSR basis as it
is by construction Lorentz-invariant, and reduces to the standard relativistic
particle hamiltonian in the large $\kappa$ limit. Indeed, using the fact that
for $P_\mu$ small compared to $\kappa$, in any DSR theory $\eta_\mu \sim P_\mu
+ O(1/\kappa)$ we have
\begin{equation}\label{a}
 \kappa\eta_4 = \kappa^2\sqrt{1 + \frac{P_0^2 -\vP}{\kappa^2}} \sim \kappa^2 + \frac12\left(P_0^2 -\vP\right) +
 O\left(\frac1{\kappa^2}\right)
\end{equation}

Then it follows from eq.~(\ref{22}) that \begin{equation}\label{aa}
    \eta_\mu = [x_\mu,
\kappa\eta_4] = [x_\mu,{\cal H}] \equiv \dot x_\mu
\end{equation} 
can be identified with four velocities $u_\mu$. The Lorentz
transformations of four velocities are then given by eq.~(\ref{23}) and are identical
with those of  Special Relativity. Moreover, since
\begin{equation}\label{b}
 u_0^2 - \vec{u}\,{}^2 \equiv {\cal C} = M^2
\end{equation}
by the standard argument the three velocity equals $v_i = u_i/u_0$ and the
speed of massless particle equals $1$.  Let me stress here once again that this
result is DSR model independent, though, of course, the relation between three
velocity of massive particles and energy they carry depends on a particular DSR
model one uses.
\newline

{\bf Exercise 12}. Compute the velocity of massless particles for DSR1 directly. Use $\kappa\eta_4$ as the hamiltonian and explicit expressions for $\eta_\mu$ as functions of energy and momenta (\ref{24}). (The answer can be found in \cite{Daszkiewicz:2003yr}.) \newline 

Thus this calculation indicates that GLAST should not see any signal of energy 
dependent speed of light, at least if it is correct  
to think of photons as of point massless classical particles, as I have implicitly assumed in the derivation above.

It should be stressed that the issue of velocity of {\em physical} particles is not completely settled on the theoretical ground, and thus any experimental input would be extremely valuable.

\subsection{Remarks on multi-particle systems}

Having obtained the one-particle phase space of DSR it is natural to try to
generalize this result to find the two- and multi-particles phase spaces. It
turns out however that such a generalization is very difficult, and in spite of
many attempts not much about multi-particles kinematics is known. On the other
hand the control over particle scattering processes is of utmost relevance in the
analysis of seemingly one of the most important windows to quantum gravity
phenomenology, provided by Ultra High Energy Cosmic Rays and possible
violations of predictions of Special Relativity in UHECR physics (see e.g.,
\cite{Amelino-Camelia:2002vw}, \cite{Amelino-Camelia:2003ex}  for more detailed discussion and the list of
relevant references.)

Ironically,  we have in our disposal the mathematical structure that seem to
provide a tool to solve multi-particle the problem directly. This structure is
co-product. \index{co-product} Recall that the co-product is a mapping from the algebra to the
tensor product
\begin{equation}\label{26}
\Delta: {\cal A} \rightarrow {\cal A}\otimes {\cal A}
\end{equation}
 and thus it provides the rule how the algebra acts on tensor products of its
representations. We know that in ordinary quantum mechanics two-particles
states are described as a tensor product of single-particle ones\footnote{In fact there is more to the description of multi-particles states than just the tensor product, namely one should impose somehow the statistics by symmetrizing or anti-symmetrizing the product. It is well known that in 4 dimensions these are the only possibilities, but the proof relies heavily on the assumption of Poincar\'e invariance. It is not known if relaxing this assumption by replacing the Poincar\'e with $\kappa$-Poincar\'e invariance can result in some other, braided statistics.}. Note that
this is a very strong physical assumption: in making it we claim that any
two-particle system is nothing but two particles in a black box, i.e., that the
particles preserve their identities even in multi-particle states. But it is
well possible that multi-particle states differ qualitatively from the
single-particle ones, for example as a result of non-local interactions. Let us,
however, assume that in also DSR to obtain the multi-particle states one should only
tensor the single-particle ones, and let us try to proceed.

In the case of classical algebras the co-product is trivial: 
$\Delta{G} = G\otimes1 + 1\otimes G$ which means that the group action on two particle
states just respects Leibnitz rule. For example the total momentum of two
particles in Special Relativity is just the sum of their momenta:
$$
\Delta(P_\mu)\, |1+2> = \Delta(P_\mu)\, |P^{(1)}> \otimes\, |P^{(2)}> =$$
\begin{equation}\label{27}
\left(P_\mu\otimes\bbbone + \bbbone\otimes P_\mu\right)\,|P^{(1)}> \otimes\,
|P^{(2)}> =\left(P^{(1)}_\mu + P^{(2)}_\mu\right)\,|P^{(1)}> \otimes\,
|P^{(2)}>
\end{equation}

 In the case
of quantum algebras the co-product is non-trivial and non-symmetric by definition (if the co-product was
symmetric we would have to do instead with just a classical Lie algebra in nonlinear disguise). This immediately leads to
the problem, as I will argue below.

Before turning to this problem let us point out yet another one, relevant for 
 DSR1 as well as for DSR2. Namely the co-product has been constructed so that
two-particle states transform as the single-particle ones (for example in
Special Relativity total momentum is Lorentz vector.) Indeed if we calculate the total energy and momentum of two-particles system using the co-product addition rule of DSR1 from \index{co-product!addition rule}

\begin{equation}\label{cpr1}
 \Delta (P_0) = \bbbone\otimes P_0 + P_0 \otimes \bbbone
\end{equation}

\begin{equation}\label{cpr2}
\Delta (P_k) = P_k\otimes {\rm e}^{-P_0/\kappa} + \bbbone \otimes P_k
\end{equation}

we find

\begin{equation}\label{cpr3}
   P_0^{1+2} = P_0^{(1)} + P_0^{(2)} , \quad P_k^{1+2} = P_k^{(1)}e^{-P^{(2)}_0/\kappa} + P_k^{(2)}.
\end{equation}
But then it follows that
total momentum must satisfy the same mass shell relation as the single particle
does.\newline

{\bf Exercise 13}. Check that $P_0^{1+2}$ and $P_k^{1+2}$ satisfy the dispersion relation of DSR1 if $P_0^{(1/2)}$, $P_k^{(1/2)}$ do.\newline

 We know however that in the case of the DSR1 we have to do
with maximal momentum for particles, of order of Planck mass. While  acceptable
for Planck scale elementary particles, this is certainly violated for
macroscopic bodies. To prove this, the reader can perform a nice quantum gravity phenomenology experiment 
 just by kicking a soccer ball! So we know that
there is an experimental proof that either our procedure of attributing
momentum to composite system by tensoring and applying co-product, or the
bicrossproduct DSR, or both are wrong.

To investigate things further let us turn to the DSR theory, which does not
suffer from the ``soccer ball problem'' namely to the classical \index{soccer ball problem in DSR} basis DSR with
standard dispersion relation ${\cal P}_0^2 - {\cal P}_i^2 = m^2$, for which 
de Sitter coordinates are given by (\ref{25}). The co-product for this basis has been calculated in
\cite{Kowalski-Glikman:2002jr} and up to the leading terms in $1/\kappa$
expansion read
  \begin{equation}\label{28}
\Delta({\cal P}_{0}) = \bbbone\otimes {\cal P}_{0} + {\cal P}_{0}\otimes
\bbbone + \frac1{\kappa}\, {\cal P}_i \otimes {\cal P}_i + \ldots
\end{equation}
\begin{equation}\label{29}
 \Delta({\cal P}_{i}) =\bbbone\otimes {\cal P}_{i} + {\cal P}_{i}\otimes
\bbbone + \frac1{\kappa}\, {\cal P}_0 \otimes {\cal P}_i + \ldots
\end{equation}
Using this we see that according to the co-product addition rule the total momentum of two-particles system is
\begin{equation}\label{30}
{\cal P}_{0}^{(1+2)} = {\cal P}_{0}^{(1)} + {\cal P}_{0}^{(2)} +
\frac1{\kappa}\,{\cal P}_{i}^{(1)}  {\cal P}_{i}^{(2)}
\end{equation}
\begin{equation}\label{31}
{\cal P}_{i}^{(1+2)} = {\cal P}_{i}^{(1)} + {\cal P}_{i}^{(2)} +
\frac1{\kappa}\,{\cal P}_{0}^{(1)}  {\cal P}_{i}^{(2)}
\end{equation}
As it stands, the formulas (\ref{30}, \ref{31}) suffer from two problems: first
of all, recalling that ${\cal P}_{\mu}$ transforms as a Lorentz vector for
single particle, these expressions look terribly non-covariant. Second, even
though (\ref{30}) is symmetric in exchanging particles labels $1
\leftrightarrow 2$, (\ref{31}) is not. How do we know which particle is first
and which is second? Let us try to resolve these puzzles in turn.

That the first puzzle is just an apparent paradox follows immediately from the
consistency of the quantum algebra. As I said above the action of boosts on
two-particle state is such that total momentum transforms exactly as the
single-particle momentum does. This is in fact the very reason of the ``soccer
ball problem'' in the  DSR1. In fact the boosts do not only act on
${\cal P}_{\mu}^{(1)}$ and ${\cal P}_{\mu}^{(2)}$ independently; they also mix
them in a special way. This feature was to be expected, since the co-product addition rule mixes single-particle states in a non-trivial way. More specifically, note that boosts must act on two-particle states
by co-product as well, therefore in order to find out how a two-particle state changes
when we boost it we must compute the commutator $[\Delta(N), \Delta({\cal
P})]$. Recall now that the co-product of boosts reads (again up to the leading
terms in $1/\kappa$ expansion)
\begin{equation}\label{32}
  \Delta
(N_i) = \bbbone\otimes N_i + N_i \otimes \bbbone - \frac{1}{\kappa}\, N_i
\otimes {\cal P}_0 - \frac{1}{\kappa}\, \epsilon_{ijk}\, M_j \otimes {\cal P}_k
\end{equation}
Using this one easily checks explicitly that
\begin{equation}\label{33}
[\Delta(N_i), \Delta({\cal P}_j)] = \delta_{ij} \Delta({\cal P}_0), \quad
[\Delta(N_i), \Delta({\cal P}_0)] = \Delta({\cal P}_i)
\end{equation}
from which it follows that ${\cal P}_{0}^{(1+2)}$ and ${\cal P}_{i}^{(1+2)}$ do
transform covariantly, as they should\footnote{This holds, of course, for a DSR theory in any basis, not just in the classical one.}. Of course equation (\ref{33}) holds to
all orders, as it just reflects the defining property of the co-product.

Let us now turn to the second puzzle, the apparent dependence of the total energy/momentum on physically arbitrary labelling of particles. Here I have much less to say, as this paradox has 
not been yet solved. One should however mention an interesting result obtained in the case
of the analogous problem in deformed, non-relativistic model. In the paper \cite{kosinski:1998} 
the authors find that even though there is an apparent asymmetry in particle labels due to the
asymmetry of the co-product, the representations with flipped labels are related to the original 
ones by unitary transformation, and are therefore physically completely equivalent. In the similar spirit
in \cite{Bonechi:qf} one uses the fact of such an equivalence in 1+1 dimensions to demand that 
the action of generators on two particles (bosonic) states is through symmetrized co-product.

During this Winter School Aurelio Grillo and Fernando Mendez produced another interesting puzzle concerning the validity of co-product based momenta addition rule. This puzzle reminds somehow the entanglement problem in quantum mechanics and it can be described as follows.

Suppose we use (\ref{cpr3}) to formulate conservation rule for two-to-two particles scattering, which would therefore take the following form

\begin{equation}\label{cpr4}
    P_0^{(1)} + P_0^{(2)} = P_0^{(3)} + P_0^{(4)} 
\end{equation}
\begin{equation}\label{cpr5}
    P_k^{(1)}e^{-P^{(2)}_0/\kappa} + P_k^{(2)} = P_k^{(1)}e^{-P^{(3)}_0/\kappa} + P_k^{(4)}.
\end{equation}
But what about all other particles  in the Universe (spectators)? \index{spectator problem in DSR} In principle, their presence would contribute non-trivially to the conservation laws (\ref{cpr4}), (\ref{cpr5}), to wit
\begin{equation}\label{cpr6}
    P_0^{(1)} + P_0^{(2)} + P_0^{(univ)} = P_0^{(3)} + P_0^{(4)} + P'_0{}^{(univ)}
\end{equation}
$$ 
    \left(P_k^{(1)}e^{-P^{(2)}_0/\kappa} + P_k^{(2)}\right)\, e^{-P_0^{(univ)}/\kappa} + P_k^{(univ)} $$\begin{equation}\label{cpr7}= \left(P_k^{(1)}e^{-P^{(3)}_0/\kappa} + P_k^{(4)}\right)\, e^{-P'_0{}^{(univ)}/\kappa} + P'_k{}^{(univ)}.
\end{equation}
In the standard, Special Relativistic case we neglect the influence of the rest of the Universe, because we believe that the processes are (at least approximately) {\em local}, but here we have non-local influence of one particle on another  all the time, independently of their separation (in the formulas (\ref{cpr6}), (\ref{cpr7}) there is no information concerning separation of particles in space and time.) Thus, the final construction of DSR theory must necessarily solve this spectator problem as well!

\section{Conclusion}

There is a growing hope that some form of DSR theory indeed describes Nature  in the kinematical regime, where the energies of the particles became close to the Planck energy scale and at the same time one could neglect local degrees of gravity, described by (still to be constructed) Quantum Theory of Gravity. This hope is based on the analogy between the ground state of 4d quantum gravity  and 3d quantum gravity, both being described by topological quantum  field theory.

As we saw, we seem to know some of the ingredients of the DSR theory, and we can even predict some (testable, in principle) DSR phenomenology. It seems however that there would be very hard to derive the complete form of DSR just from the first principles, the hope being that soon we will be able to derive DSR as an appropriate limit of (Loop) Quantum Gravity.

Four years ago during the Winter School entitled ``Towards Quantum Gravity''  Giovanni Amelino-Camelia asked the insightful question ``Are we at the dawn of Quantum Gravity Phenomenology?''. This year we devoted the whole Winter School to discuss possible observable signals of Quantum Gravity. I hope that in four years we will meet to discuss numbers coming from Quantum Gravity experiments that would be already running and producing data. I also hope that it will turn out that these data would agree with the final form of Doubly Special Relativity.

%\printindex
\end{document}